\documentclass{article}
\usepackage{dafne99pro}
\begin{document}
\title{A NEW MEASUREMENT OF THE MUON MAGNETIC ANOMALY}
\author{ 
\hspace*{5mm} K.Jungmann$^f$, H.N.Brown$^b$, G.Bunce$^b$, R.M.Carey$^a$, P.Cushman$^j$,\\
 G.T.Danby$^b$, P.T.Debevec$^h$, H.Deng$^l$, W.Denninger$^h$, S.K.Dhawan$^l$\\
 V.P.Druzhinin$^c$, L.Duong$^i$, W.Earle$^a$, E.Efstathiadis$^a$, F.J.M. Farley$^l$\\
 G.V.Fedotovich$^c$, S.Giron$^j$, F.Gray$^h$, M.Grosse-Perdekamp$^l$, A.Grossmann$^f$\\
 U. Haeberlen$^g$, E.S.Hazen$^a$, D.W.Hertzog$^h$, V.W.Hughes $^l$, M.Iwasaki$^k$\\
 D.Kawall$^l$,  M.Kawamura$^k$, B.I.Khazin$^c$, J.Kindem$^j$, F.Krienen$^a$, I.Kronkvist$^j$\\
 R.Larsen$^b$, Y.Y.Lee$^b$, I.Logashenko$^a$, R.McNabb$^j$, W.Meng$^b$, J-L.Mi$^b$\\
 J.P.Miller$^a$, W.M.Morse$^b$, D.Nikas$^b$, C.Onderwater$^h$, Y.Orlov$^d$, C.Ozben$^b$\\
 C.Pai$^b$, J.Paley$^a$, C.Polly$^h$, J.Pretz$^l$, R.Prigl$^b$, G.zu Putlitz$^f$, S.I.Redin$^l$\\
 O.Rind$^a$, B.L.Roberts$^a$, N.M.Ryskulov$^c$, S.Sedykh$^h$, Y.K.Semertzidis$^b$\\
 Y.Shatunov$^c$, E.Solodov$^c$, M. Sossong$^h$, A.Steinmetz$^l$,  L.R.Sulak$^a$\\
 C.Timmermans$^j$, A.Trofimov$^a$, D.Urner$^h$, P. V. Walter$^f$, D.Warburton$^b$\\
 \hspace*{1.5cm}D.Winn$^e$, A.Yamamoto$^i$, D.Zimmerman$^j$\\
\newline
\em
$^a$Boston University;
$^b$Brookhaven National Laboratory;
$^c$Budker Institute of\\
Nuclear Physics, Novosibirsk;
$^d$Cornell University;
$^e$Fairfield University;\\
$^f$University of Heidelberg;
$^g$MPI f. Med. Forsch., Heidelberg;
$^h$University of \\
Illinois;
$^i$KEK;
$^j$University of Minnesota;
$^k$Tokyo Institute of Technology;\\
$^l$Yale University      
 }
\maketitle
\baselineskip=11.6pt
\begin{abstract}
The muon magnetic anomaly may contain contributions from physics beyond the 
standard model.
At the Brookhaven National Laboratory (BNL) a precision experiment aims
for a measurement of the muon magnetic anomaly $a_{\mu}$  to 0.35 ppm,
where conclusions about various theoretical approaches beyond
standard theory can be expected.
The difference between the spin precession and cyclotron frequencies
is measured in a magnetic storage ring with highly homogeneous field. 
Data taking  is in progress and part of all recorded data 
has been analyzed. Combining all experimental results to date 
yields preliminarily $a_{\mu}(expt)=1~165~921(5) \cdot 10^{-9}$ (4 ppm) in
agreement with standard theory. 
 
\end{abstract}
\baselineskip=14pt

\section{Physics Motivation}

The magnetic anomaly of fermions $a=\frac{1}{2}\cdot(g-2)$
describes the deviation of their magnetic g-factor
from the value 2 predicted in the Dirac theory. 
This quantity 
has been measured for single electrons and positrons in Penning traps
by Dehmelt and his coworkers to 10~ppb \cite{Dyc_90}. 
Accurate calculations for $a$ of these two particles are possible to this level,
which involve exclusively the "pure" Quantum Electrodynamics (QED)
of electron, positron and photon fields.
The presently most accurate 
value for the fine structure constant ¼$\alpha$ \cite{Hug_99}
can be obtained from a comparison between experiment and theory, 
where it appears as an expansion coefficient.
The high accuracy, to which QED calculations 
can be performed, is demonstrated by the 
compatibility of this value of $\alpha$ and the ones obtained in
measurements based on the quantum Hall effect \cite{QH} 
or the number extracted from the 
precisely known Rydberg constant 
using an accurate determination of the neutron de Broglie wavelength 
and relevant mass ratios. 
Moreover, the agreement of $\alpha$ values determined 
from the electron magnetic anomaly and from the hyperfine splitting in 
the muonium atom \cite{Liu_99}
may be interpreted as the most precise 
reassurance of the internal consistency of QED, 
because the first case involves the theory of 
free particles  whereas in the second case distinctively 
different bound state approaches need to be applied \cite{Kin_90}. 

The anomalous magnetic moment of the muon $a_{\mu}$ is more sensitive
by a factor of $(m_{\mu}/m_e)^2 \approx 4 \cdot 10^{4}$ 
to heavier particles, which 
appear virtually in loop graphs,
and other than electromagnetic interactions.
Such effects can be studied in a precise determination of $a_{\mu}$, 
because very high confidence in the validity 
of calculations of the dominating QED contribution arises from the excellent
description of the electron magnetic anomaly and electromagnetic 
transitions in fundamental systems like, e.g.  hydrogen and  muonium atoms\cite{Bos_96}. 

In a series of three experiments at CERN \cite{cern}
$a_{\mu}$ could be  measured to 7.2~ppm.
This has verified the muons nature as a heavy leptonic particle and
the proper description of its electromagnetic interactions to very high accuracy
by QED.
In the last of these measurements contributions arising from strong 
interactions, 
which amount to 57.8(7)~ppm \cite{Dav_98}, could be verified.
At BNL a new dedicated experiment has been designed
to determine the muon magnetic anomaly $a_{\mu}$ with 0.35 ppm relative accuracy, 
meaning a 20 fold improvement over the previous approaches. 
At this level exists particular sensitivity to contributions 
arising from weak interaction through loop diagrams with W and Z bosons (1.3~ppm)
\cite{Cza_96}. 
The experiment promises here a clean test of renormalization in weak interaction.
The muon magnetic anomaly may  also contain
contributions from new physics \cite{Mer_90}.
A variety of speculative theories can be tested which have been invented to
extend the present standard model in order to explain some of the features
which are described but not fundamentally understood yet.
The spectrum of such theoretical models 
includes physics concepts like muon substructure, new gauge bosons,
supersymmetry, an anomalous magnetic moment of the W boson, leptoquarks
and violation of Lorentz and CPT invariance.
Here a precise measurement of $a_{\mu}$ can be  complementary 
to searches
in high energy experiments and the sensitivity 
may even be higher.

\section{The Brookhaven Muon g-2 Experiment}

In the new experiment at the alternating gradient synchrotron (AGS) of BNL
polarized muons are stored  in a magnetic storage ring
with highly homogeneous field $B$ and with weak electrostatic focussing.
The difference §$\omega_a$
of the spin precession and the 
cyclotron frequencies,
\hspace{1mm} 
\begin{equation}
\omega_a =  \omega_s - \omega_c = a_{\mu} \frac{e}{m_{\mu}c}B  ,
\end{equation}
is measured,
with $m_{\mu}$ the muon mass and $c$ the speed of light.
Positrons (electrons)  from the weak decays
$\mu^{\pm}\rightarrow e^{\pm} + 2\nu$ are observed.
For relativistic muons the influence of a static electric field vanishes 
\cite{Telegdi}, if §$a_{\mu} = 1/( \gamma_{\mu}^2-1)$ 
which corresponds to $\gamma_{\mu}=29.3$ and a muon momentum of $p=3.09$ GeV/c,
where $\gamma_{\mu}= 1/ \sqrt{1-(v_{\mu}/c)^2}$ and $v_{\mu}$ is
the muon velocity.
For sufficient accuracy of the electric field correction the average
muon momentum $p$ needs to be within a few parts in 10$^{4}$ of magic momentum.

For a homogeneous field the magnet must have iron flux return and shielding.
Because of the particular momentum requirement and in order to
avoid strong magnetic saturation effects of the iron a device of 
7~m radius was built. It has a C-shaped iron yoke cross section with 
the open side facing towards the center of the ring.
It provides 1.4513~T field  in a 18~cm gap. The magnet is energized by 4
superconducting coils carrying 5177~A current.

\begin{figure}[t]
 \vspace{9.0cm}
\includegraphics{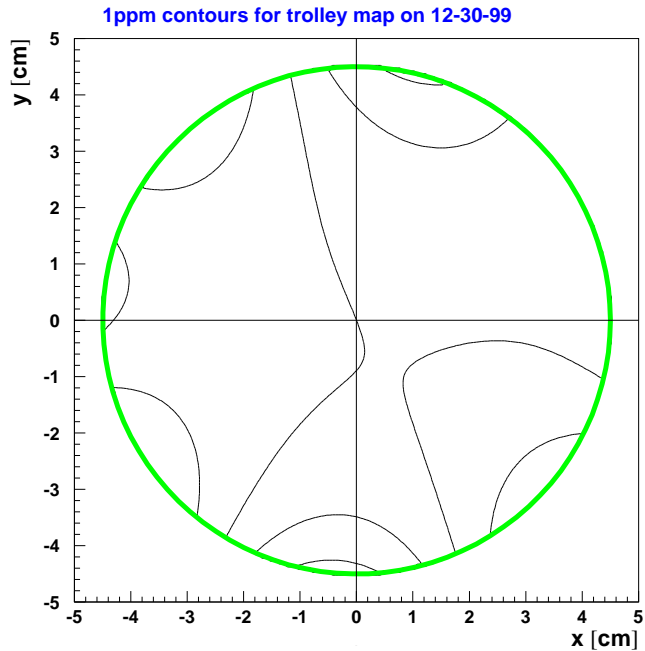}
 \caption{\it
  Contours of the average  magnetic field around the ring 
  (1 ppm between lines).
  for one arbitrary  measurement with the beam tube trolley. 
   The circle line indicates the storage volume boundary.
    \label{exfig1} }
\end{figure}

The magnetic field is determined by a newly developed 
narrow band magnetometer system
which is based on pulsed nuclear magnetic resonance (NMR) of protons in water
and vaseline.
It has a capability for an absolute measurement to $\approx 50$~ppb.
\cite{Pri_96}. The field and its homogeneity are  continuously monitored by 
380 NMR probes. They are  distributed around the ring
and they are embedded near the magnet poles in the walls of the Al vacuum 
tank.  
For mapping the field inside the storage volume a trolley 
carrying 17 NMR probes
is run in regular intervals, typically twice a week. 
This device contains a fully computerized magnetometer built entirely from
nonferromagnetic components.
The field accuracy is derived from and related
to a precision measurement of the proton gyromagnetic ratio in 
a spherical water sample  \cite{Phillips_77}.
On average the field around the ring is homogeneous to 1~ppm (Fig.\ref{exfig1}).
This has been achieved  using mechanical shimming methods which include
movable iron wedges in an air gap between the low carbon steel
pole pieces and the magnet yoke as well as iron strips of adjusted width fixed to
the neighbourhood of junctions between poles. A set of 60 electrical coils, 
which run on the surface of the pole pieces around the ring and which 
can be driven at individually different currents, 
 allows the compensation of other than dipole components of the field.
The absolute value of the field integral in the storage region is 
known at present to better than 0.5 ppm. 
There is a potential for a significant improvement in this figure. 
Field drifts  are compensated using a set of 36 selected fixed NMR probes. 
Their average is kept within 0.1 ppm of the nominal value 
by regulating the main magnet power supply. 
To avoid large short term thermal effects, the magnet yoke has been  dressed with 
passive thermal insulation material.

The weak muon focussing is provided by  
electrostatic quadrupole electrodes with 10~cm 
separation between opposite plates. 
They cover 43 \% of the ring circumference.
The electric field is applied by pulsing
$\pm 24.5$~kV voltage for $1.4$ms  duration 
to minimize electron trapping and avoid electrical breakdown.
The storage volume diameter is defined by circular apertures to 9~cm.

\begin{figure}[t]
 \vspace{9.0cm}
\includegraphics{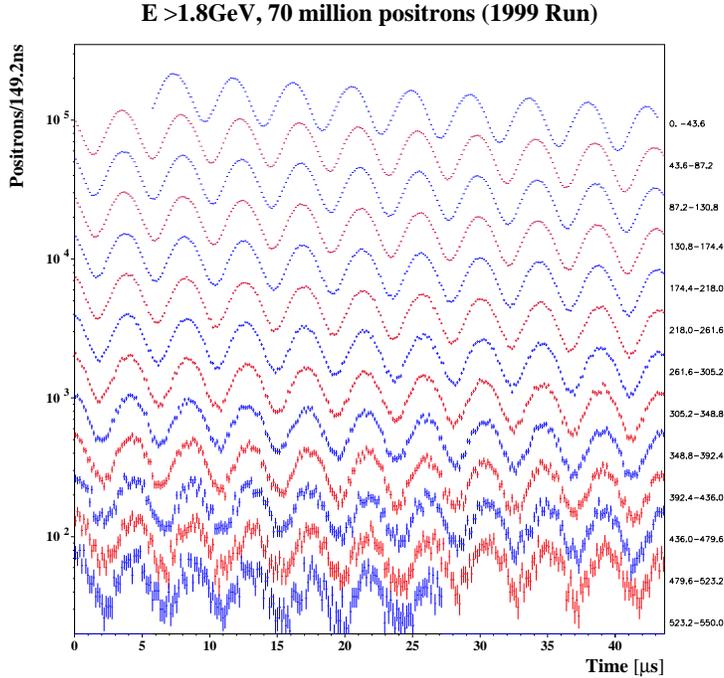}
 \caption{\it
   A sample of the recorded data. Positrons with an energy exceeding 
   1.8GeV as counted in the detector stations as a function of time
   after injection.  Several regions in time have been 
   folded onto one graph with the boundaries (in $\mu$s) mentioned 
   to the right in each case . 
   Time dilatation results in a 64.4  $\mu$s lifetime. No evidence for significant background
   is visible after 8 lifetimes. 
    \label{exfig2} }
\end{figure}

Due to parity violation in the weak muon decay process 
the positrons (electrons) are emitted preferentially
along (opposite) to the muon spin direction. This causes a time dependent
variation of the spatial distribution of decay particles in the muon 
eigensystem which translates into a time dependent variation of the energy
distribution in this experiment.
Inside the ring the positrons(electrons) 
are observed in 24 shower detectors consisting of
scintillating fibers embedded in lead.
They have 13 readiation lengths thickness and an average resolution of $\sigma$/E = 6.8\% 
at the nominal energy cut of E = 1.8 GeV which is 
applied in the analysis leading to the positron distribution shown in
Fig.  \ref{exfig2}.
All positron events 
are digitized individually in a custom waveform digitizer
at 400 MHz rate and stored for analysis. The time standard of the detectors
and the field measurement system is a single LORAN C receiver with 
better than $10^{-11}$ long term stability.

The technical improvements over previous experiments at CERN include
an azimuthally symmetric iron construction for the magnet with 
superconducting coils,
a larger gap and higher homogeneity of the field,
segmented positron (electron) detectors
covering a larger solid angle and improved electronics.
A major advantage is the two orders of magnitude  
higher primary proton intensity
available at the AGS Booster at BNL. Further conceptually novel  features 
are the NMR trolley, a superconducting static inflector magnet and   
direct muon injection onto storage orbits by means of a magnetic kicker.
Previously pions had been introduced in to the ring
some of which decayed into stored muons. The electrostatic quadrupoles at BNL have twice
the field gradient compared to the CERN experiment and in addition the vacuum
requirements are more relaxed due to a new design which minimizes electron
trapping. The vacuum chamber is scalloped to avoid preshowering.

\section{Present Status of results}

By now data taking has been carried out for $\mu^+$ 
in two extended periods. 
In the startup phase of the experiment in 1997 \cite{carey} 
pion injection was used. The efficiency of this process was  below the 
theoretical expectation, which is $25 \cdot 10^{-6}$,
resulting in $\approx 10^3$ stored  muons per injection pulse 
(with $5 \cdot 10^{12}$ protons from the AGS on target). 
This method is accompanied by 
a significant flash in the detectors caused by hadronic interactions of unused pions.
The impact of this effect was minimized by gated photomultiplier operation.
The data were useful only after 22-75~$\mu$s, depending on the detector position.
The first result obtained in this way was 
$a_{\mu^{+}} = 1~165~925(15) \cdot 10^{-9}$(13 ppm)\cite{carey}. 
  
\begin{figure}[t]                           
  \vspace{9.0cm}
  \includegraphics{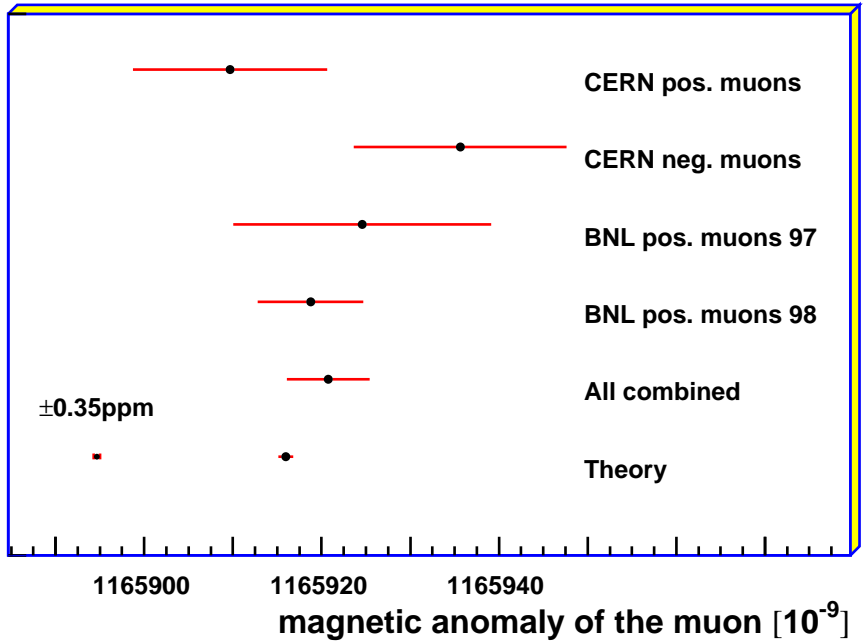}
  \caption{\it
  Results of the last CERN experiments for both muon charge states, the result
  from the startup run using pion injection in this experiment, the 
      magnetic anomaly extracted from 4\% of the recorded data and the combined
  result of all which has 4~ppm uncertainty. The experimental goal is 
  a one order of magnitude higher precision.
  \label{exfig3} }
\end{figure}

Muon injection, which is employed regularly since 1998 with an efficiency of order
5\%, gives about an order of magnitude more muons per injection pulse
and largely reduces flash background. 
In addition, major improvements in the magnetic
field homogeneity and stability were made and the 
detector efficiency was increased. 
A part of the new data ($\approx$ 4\%) have already been
completely analyzed and  provides the preliminary value of 
$a_{\mu^{+}} = 1~165~919(6) \cdot 10^{-9}$ (5~ppm) 
where the uncertainty is  dominated by  statistics.  
Among the systematic errors the dominating contributions arise 
from positron pileup in the detectors, 
flashlets, i.e. the additional delivery of small bunches of protons after
the AGS main pulse, and the field calibration. 
Combining all the measured values from CERN and BNL  (Fig.\ref{exfig3}) yields
$a_{\mu}(expt)=1~165~921(5) \cdot 10^{-9}$ (4 ppm). 
This agrees with the latest theoretical value \cite{Hug_99} $a_{\mu}(theor) =
116~591~628(77) \cdot 10^{-11}$ (0.66 ppm). The dominating error here arises
from the knowledge of the hadronic part, which has been calculated using
electron positron annihilation into hadrons and hadronic 
$\tau$-decays \cite{Dav_98}.

\section{Perspectives}

The data recorded up to now cover more than $2 \cdot 10^9$ decay positrons.
This leads to an expected  statistical uncertainty at the 1~ppm level.
Further data taking is in progress, now with typically $40 \cdot 10^{12}$
protons per AGS cycle which provides 10 pulses. 
The systematic errors are expected to sum
up to a few 0.1~ppm. In order for the new muon g-2 experiment to reach its 
0.35~ppm
design  accuracy,  besides $\omega_a$ and the field 
also the muon mass respectively its magnetic moment 
needs to be known to 0.1~ppm or better (see eq.(1)).
This has been achieved very recently
by microwave spectroscopy  of the Zeeman effect in the muonium 
atom ($\mu^+e^-$)
ground state hyperfine structure, resulting in a measurement of the ratio of 
the muon magnetic moment 
to the proton magnetic moment $\mu_{\mu}/\mu_p$ to 120~ppb.
(A comparison of the simultaneously obtained muonium ground hyperfine interval
with QED theory may be interpreted in terms of an even more precise 
value of this quantity at 30~ppb.)

In minimal supersymmetric models, as a particular example of relevant speculative models,
a contribution to $a_{\mu}$ of
\begin{equation}
  \Delta a_{\mu}(SUSY) / a_{\mu}
   \approx 1.25~{\rm ppm} \left(\frac{100\, GeV/c^2}{\tilde{m}}\right)^2 \cdot \tan \beta,
\end{equation}
is expected, 
where $\tilde{m}$ is the mass of the lightest supersymmetric particle and
$\tan \beta$ is the ratio of the vacuum expectation values for the 
two involved Higgs fields. At the projected accuracy for g-2,
there is a sensitivity to large values of the latter parameter.      

The experiment is planned for both $\mu^+$ and  $\mu^-$ as a test of CPT invariance.
There is actual interest in view of the suggestion
\cite{
Deh_99}  
to compare tests of CPT invariance in different systems on a common basis, i.e.
by using the energies of the states involved. 
For fermion magnetic anomalies particles with spin down in an external 
field need to be compared to their antiparticles with spin up.
The nature of g-2 experiments is such that they provide a figure of merit  
$r = |a^- - a^+| \cdot \frac{\hbar\omega_c}{m \cdot c^2}$ for a CPT test, 
where $a^-$ and $a^+$ are respective magnetic anomalies, and $m$ is the particle mass.
For the past electron and positron measurements one has 
$r_e \leq 1.2 \cdot 10^{-21}$ \cite{Deh_99}
which is a much tighter bound than from the neutral kaon system,
were the mass differences between $K^0$ and $\overline{K^0}$ yield 
$r_{K} \leq 1\cdot 10^{-18}$.
An even more stringent CPT test arises therefore 
already from the past muon magnetic anomaly  measurements
were  $r_{\mu} \leq 3.5 \cdot 10^{-24}$. Hence, this may be viewed as
the presently best known CPT test based on system energies.
The BNL g-2 experiment allows to
look forward to a 20 times more precise test of this fundamental symmetry.

According to the standard theory an elementary particle is not allowed to have a
finite permanent electric dipole moment (edm) as this would violate CP 
and T symmetries,
if CPT is assumed to be conserved.
An edm of the muon would manifest itself in the  g-2 experiment in a
time dependent up down asymmetry of decay positrons which can be searched for
along with the muon g-2 measurements. The BNL experiment
is expected to provide one order of magnitude  improvement
over the present limit at $1.05 \cdot 10^{-18}$ e~cm. This is possible through
proper segmentation of the detector packages.
A further highly promising approach has been suggested as a dedicated follow on
experiment. It is expected that based on the g-2 setup an experiment 
can be tailored to allow
5-6 orders of magnitude increase in sensitivity. 
It should be noted that a non standard model value of $a_{\mu}$ would call
for a muon edm search, as both quantities are intimately linked
in many theories, where their sizes are connected through a CP violating
phase  \cite{Yannis_97}. Another possibility is using the magnet
as spectrometer in which pion decays are  observed for restricting the muon
neutrino mass by a further factor of 20.

\section{Acknowledgements}
This work was supported in part by the U.S.
Department of Energy, the U.S. National Science Foundation, the German
Bundesminster f\"{u}r Bildung und Forschung, the Russian Ministry of Science
and the US-Japan Agreement in High Energy Physics.


%
\end{document}